# Metastable magnetic bubble in [Co/Pd]$_4$/Py multilayers

Yurui Wei[1], Chengkun Song[1], Yunxu Ma[1], Hongmei Feng[1], Chenbo Zhao[1], Xiaolei Li[1], Chendong Jin[1], Jinshuai Wang[1], Chunlei Zhang[1], Jianbo Wang[1,2], Jiangwei Cao[1] and Qingfang Liu[1*]

1) *Key Laboratory for Magnetism and Magnetic Materials of the Ministry of Education, Lanzhou University, Lanzhou 730000, People's Republic of China.*
2) *Key Laboratory for Special Function Materials and Structural Design of the Ministry of the Education, Lanzhou University, Lanzhou 730000, People's Republic of China.*

Corresponding author: Qingfang Liu
E-mail address: liuqf@lzu.edu.cn

**Abstract**

Magnetic bubbles are topologically spin textures that offering the interesting physics and great promise for next-generation information storage technologies. The main obstacles so far are that magnetic bubbles are generated with no field stimuli in new material systems at room temperature. Here, we report the observation of individual magnetic bubbles and its high frequency measurement at room temperature in an exchange-coupled [Co/Pd]$_4$/Py multilayers. We demonstrate that the emergence of magnetic bubbles at remanence can be tuned by the in-plane tilted magnetic field (roughly 3°) along the film plane at room temperature. High frequency results indicate that the presence of magnetic bubbles leads to broadening of the magnetic permeability spectrum lines (due to the non-uniformity of the magnetic moments). Our findings open the door to the bubble-based spintronics at room temperature in exchange-coupled magnetic multilayers.

## 1.Introduction

Metastable states [1-3], the non-equilibrium state of free energy when the system is above equilibrium states, are vital for the materials application since some properties of the metastable material outperforms its equilibrium properties or even exhibit special properties. Therefore, the study of the metastable states of materials not only exhibits the theoretical significance, but also has the important practical value. A vortex-like magnetization texture, called skyrmion, is one of the typical metastable state under room temperature when it was first found in non-centrosymmetric bulk materials with B20 structure [4-7]. Its ancestor is called magnetic bubble, which exist in ferromagnetic compounds such as garnet film [8] and alloy films [9-10] with perpendicular anisotropy. A magnetic bubble is a rod-like domain with out-of-plane magnetization embedded in a thin plate with antiparallel magnetization background and separated with a cylindrical Bloch wall. Magnetic bubbles and skyrmions are closely relative as they exhibit in the same topology [11-12]. However, their characteristic sizes are different [13-17]. Since magnetic bubbles present a long lifetime at room temperature, which made them attracts attention in recent years. Various imaging techniques, such as Scanning Probe Microscopy [17], Lorentz transmission electron microscopy [14-15], and magnetic-force microscopy [6,9], et al [13,18-20], were used to the detection of magnetic bubbles in ultrathin films.

The formation of magnetic bubble depends on some factors [9,26], such as the quality factor $Q = K_u / 2\pi M_s^2$ (the ratio of the perpendicular magnetic anisotropy and magnetostatic energies). If Q>1, the stripe domains are formed in magnetic films. With applied magnetic field perpendicular to the film surface, the domain states transform to bubble domain states and the bubble size strongly depends on the external field [21-23]. As an intermediate state, the stripe domain is important for formation of magnetic bubble. However, in some alloy single-layer film (like permalloy film), the strong interfacial perpendicular anisotropy is not easy to obtain and the stripe domain exists above the critical thickness (for Py film [24], the critical thickness is larger than 80 nm), which prevents the existence of magnetic bubble in this kind of film under the critic thickness.

In this work, we fabricate [Co/Pd]$_4$/Py multilayers, and obtained the stripe domain state in Py layer with 15 nm. After magnetized by an in-plane titled magnetic field then measured at remnant state, the stripe domain state changed to magnetic bubble. The stability and chirality as well as the high-frequency properties of these magnetic bubbles are then investigated with magnetic force microscope (MFM) and vector network analyzer ferromagnetic resonance (VNA-FMR) techniques.

**2. Experimental methods**

The magnetic multilayers with the composition of Ta(10 nm)/Py(15 nm)/Ta(3 nm), Ta(10 nm)/Pd (5 nm)/[Co(0.3 nm)/Pd(0.8 nm)]$_4$/Ta(3 nm) and Ta(10 nm)/Pd (5 nm)/[Co(0.3 nm)/Pd(0.8 nm)]$_4$ / Py(15 nm)/Ta(3 nm), where Py denotes permalloy (i.e., Ni$_{81}$Fe$_{19}$), were deposited onto a thermal oxidized Si substrate by magnetron sputtering with the base pressure below $1.9\times10^{-7}$ Torr [Figure 1(a)]. The Pd and Ta layers were deposited by radio-frequency (RF) sputtering, while the Py and Co layers were deposited by DC magnetron sputtering (the sputtering pressure of Ar gas was kept at 5 mTorr for all layers). A 3-nm Ta cap was deposited to prevent oxidation of the stack. The thin amorphous Ta (10 nm) seed layer allows for greater mobility of the deposited atoms and an improved fcc-(111) orientation of the Pd layer deposition, thus improving the perpendicular anisotropy of Co/Pd multilayers [26].

M–H hysteresis loops were measured using vibrating sample magnetometry (VSM, ADE technologies, EV9) and focused magneto-optical Kerr microscopy (NanoMOKE2). Magnetic domain structure was observed by a commercial MFP-3D AFM/MFM (Asylum Research). The MFM tips (ASYHMFM) were ~84 nm in radius and kept at a distance h of 20 nm from the sample surface. Ferromagnetic resonance measurements (FMR) were fulfilled using vector network analyzer (VNA) as a function of the applied microwave magnetic field under different dc magnetic field. All measurements were performed at room temperature.

**3. Results and discussion**

Figure 1(c) shows the hysteresis loops of the Ta(10 nm)/Pd(5 nm)/[Co(0.3 nm)/Pd(0.8 nm)]$_4$/ Py(15 nm)/Ta(3 nm) film with the applied field perpendicular and parallel to the film plane. For comparison, the result of film without Py layer (that is Ta(10 nm)/Pd(5 nm)/[Co(0.3 nm)/Pd(0.8 nm)]$_4$/Ta(3 nm)) is also shown in Figure 1(b). The hysteresis loop of [Co/Pd]$_4$ multilayer shows obvious out-of-plane anisotropy with the effective anisotropy constant $K= 8.1 \times 10^5$ J/m$^3$, which is smaller than that obtained in reference 25. However, the out-of-plane hysteresis loop of [Co/Pd]$_4$/Py exhibits a two-steps reversal, because the film is composed of a soft magnetic layer Py and exchange-coupled (Co/Pd)$_4$ magnetic layers which the soft magnetic layer is first reversed and then the exchange-coupled layer flipped. Due to the existence of Py, the magnetic moment is easy to tilt an angle from the out-of-plane direction, which is due to the competition between the in-plane anisotropy of Py and the out-of plane perpendicular anisotropy of (Co/Pd)$_4$. That leads to a lower saturated field in in-plane hysteresis loops of [Co/Pd]$_4$/Py film. In addition, the in-plane hysteresis loop of [Co/Pd]$_4$/Py film shows a little horizontal shift along the magnetic field with a bias field of about 4 Oe, which may due to the formation of un-equally oriented closure domains [27] or vortex cores [28] at the [Co/Pt]$_n$/NiFe interface after in-plane saturation. If the out-of-plane anisotropy increase (realized by increase of stack number n in (Co/Pd)$_n$ to n=6 in our experiment), the exchange-bias became more obvious. The MOKE loop of Py layer [Figure 1(d) and 1(e)] in our [Co/Pd]$_4$/Py film shows that the easy magnetized axis is still lying in plane, although the out-of-plane MOKE hysteresis loops show a high remanence ratio, indicating perpendicular component of magnetic moment in Py layer due to the strong magnetostatic and exchange interaction between Py and (Co/Pd)$_4$ layer.

Figure 2(a)-2(d) show the image of magnetic domain of our [Co/Pd]$_4$/Py film, measured by MFM under a vertical magnetic field to the film plane at room temperature. In Figure 2(a), we present the MFM imaging at zero fields. One can notice that micron-sized domains exist in the image. After the magnetic vertical field B=10 mT is applied [Figure 2(b)], the stripe domains with magnetization perpendicular to the film plane grow at the expense of domains opposite to the film plane. As the vertical field increasing to $B$=18 mT, the labyrinth domains with longer sizes are observed. When the field increase to 24 mT, all the stripes and labyrinth domains completely transformed into random magnetic bubble with the average diameter of about 420 nm, as shown in Figure 2(d). When the vertical field exceed 25 mT, the magnetic bubble annihilated and the film was saturated to a ferromagnetic state.

The above results show that the labyrinth domain is sensitive to magnetic field. Next, we changed the magnetized method and analysis the domain structure of our (Co/Pd)$_4$/Py film again. Before MFM observation, the sample was magnetized under an in-plane tilted field (roughly 3° to the plane as shown in Figure 3) with a small out-of plane component, then the magnetic field is removed, the whole process spend about 30 seconds. Figure 3(a)-(i) show the MFM images after magnetized under different tilted field (from 0 mT to 1000 mT). Each MFM image was obtained from the same region of the sample, but not the same specific area of the sample surface. Figure 3(a) shows the snap-shot of the MFM imaging at zero fields. One can notice that the stripe domains, the micron-sized domains (large white areas), and magnetic bubbles coexist in the image. The images reveal that the out-of-plane magnetization orientation averaged over the thickness of the magnetic domains through the variations of dark/bright intensity contrast. After magnetized under 100-500 mT [Figure 3(b)-(d)], the larger white domain disappeared due to the perpendicular component of tilted field, only left the labyrinth domain with a size of 2 μm. At the magnetized field of 800 mT [Figure 3(e)], all labyrinth domains completely transformed into disordered magnetic bubbles (white) with the diameter about 580 nm.The reason we didn't show the labyrinth patterns is because the maze domain still occurs in the range of 500–800 mT . As the magnetized field further increase, the size and number of bubbles gradually becomes smaller. At 950 mT, the white bubbles have the smallest size of ~230 nm [Figure 3(f)]. After the magnetized field increase to 1000 mT, the bubbles completely disappear [Figure 3(i)]. During our experiment, we also observed the black bubbles [Figure 3(g), 900 mT] if we change the magnetized field to -3° direction to the film plane (-z component field) and remove it before MFM observation. Further increase of magnetized filed leads to the disappearance of some bubbles as shown in Figure 3(h). This result indicates that the tilted field with small vertical magnetic field component can indeed trigger bubbles in opposite polarity[20].

The above results show that after magnetized at certain tilted magnetic field, magnetic bubbles state in Py layer is triggered. To verify the thermal stability [29-32], we excited the magnetic bubble state at the tilted field of 650 mT then removed the magnetic field. At the remnant state, the MFM was used to observe the film again. After first observation, the sample is not taken away from the holder to ensure the next observation in the same area after waiting different time (1, 3, 5, 7, 9 hours). The domain structure under different relaxation time is shown in Figure 4(a-f). No changes in the domain structure were observed after 9 hours, indicating the bubble state is stable at zero fields within 9 hours. However, after three days, the bubble state disappears (not show). It means that the magnetic bubble state in our (Co/Pd)$_4$/Py film is a metastable state. In addition, the labyrinth magnetic domains in our film may have chirality because we obtained the non-zero DMI coefficient D=0.38 mJ/m$^2$ in our (Co/Pd)$_4$/Py film by using momentum-resolved BLS measurements. Since the thickness of Py is 15 nm, we think that the interfacial DMI mainly comes from the interface of Co/Pd [seen at Supplementary Material S1] .

To figure out the chirality of our magnetic bubble, micromagnetic simulations were performed with the MuMax3 code. The simulated model includes a Py layer placed at the top of [Co/Pd]$_4$ layers. The dimension of the sample is 1200 nm×1200 nm×20 nm, in which the NiFe film is 16 nm thick and Co is 4 nm thick. We take the following material parameters for the soft NiFe and hard [Co/Pd]$_4$ layers [25-26]: exchange constant $A_{Py} = 13 \times 10^{-12} J/m$, $A_{Co/Pd} = 6 \times 10^{-12} J/m$ . The perpendicular magnetic anisotropy constant of [Co/Pd], $K_{Co/Pd} = 8.1 \times 10^5 J/m^3$ and saturation magnetization $M_{Py} = 8 \times 10^5 A/m$ , $M_{Co/Pd} = 5 \times 10^5 A/m$ were obtained from VSM measurement. To ensure the labyrinth domain in Py layer, an effective non-zero perpendicular magnetic anisotropy $K$ is taken $\left(2-3\times10^5 J/m^3\right)$ which originates from the dipolar interaction from the Co/Pd layers. In the simulation, we have to approximate the results of the phenomenological model. The general trend can be seen although the difference with the real system. The system is discretized into a mesh of dimensions 4 nm×4 nm×4 nm. The damping parameter is set as $\alpha = 0.3$ . Due to the non-zero DMI coefficient in our multilayers, the DM interaction is considering in the Co/Pd layers.

Followed the experimental result, we first simulate the field-driven evolution of labyrinth domain, as shown in Figure 5(a)-(h), to obtain the magnetic bubble state. At 200 mT vertical field along +z direction, most labyrinth domain changed to bubble. The random isolated bubble state absolutely emerged at the magnetic field of 250 mT. By increasing the magnetic field further to 300 mT, the size of bubble is reduced [Figure 5(g)]. If the magnetic field keeps increasing further, the bubble annihilated gradually and the film was finally saturated to a ferromagnetic state.

By analyzing the magnetic moment distribution, we find there exists magnetic bubbles with topological number 1, and -1, which is labeled by colour circle shown in figure 5(g). The enlarged images of these skyrmion are also shown in figure 5(h). The existence of these bubbles with topological number of 1 and -1 in Py layer may due to the influence of Co/Pd layer with a non-zero DM interaction, which prefer to host a Neel skyrmion. To verify this fact, the domain structure of Co layer is analyzed, which shows that, under the vertical magnetic field, there exists Neel skyrmion. Due to the exchange interaction between Co and Py at the interface, the Neel skyrmion in Co layer changes to Bloch-like skyrmion (or bubble) in Py layer by orienting the magnetic moment out of the radial direction. Figure 5(i) shows the magnetization distributions of several bubbles in upper Py , interface Py and interface Co layers. One can see that the magnetic moment transitional process of the bubble from Co to Py. We can see Neel skyrmions corresponding to the interface of Py and the Co layer. From interface Py to upper Py, Bloch skyrmions occurs in the picture. Our simulation results confirm that bubble domains appearing in Co/Pd multilayer penetrate through the whole multilayer system.

Recently, understanding of the dynamics of magnetic bubble with non-zero topological number (skyrmions) under external fields is an important issue for their manipulation[33-35]. Besides the lowest energy translation mode [34], breathing, clockwise and anticlockwise radial modes have already been investigated [35]. However, due to the sensitivity of signal in ultrathin film with perpendicular magnetic anisotropy, the microwave magnetic property is difficult to obtain by vector network analyzer ferromagnetic resonance (VNA-FMR). In our [Co/Pd]$_4$/Py film, Py has a thickness of 15 nm, which makes it possible to measure the high-frequency properties of magnetic bubble in Py layer.

Figure 6 shows the imaginary part of the permeability spectra of our [Co/Pd]$_4$/Py film under different magnetic states. During the measurement, the static field and microwave field are applied in the plane as indicated in Figure 6(a) and 6(b). Figure 6(a) shows the imaginary part of permeability spectra for the bubble state, which is obtained by magnetized the sample under titled field of 800 mT and then removed the field. During the measurement of permeability, the in-plane field is in the range of 0 to 200 Oe to ensure the bubble state does not destroyed by in-plane field. As comparison, no magnetic bubble state (after magnetized the sample under 1000 mT titled field) was also measured. The results are shown in Figure 6(b). It can be seen that, for no bubble state, a sharp resonance peaks occurs at near 1 GHz under zero magnetic field, which is in accordance with the in-plane uniform precession mode. As the magnetic field gradually increases, the resonance peak moves to the higher frequency. At 200 Oe field, the maximum frequency of 4.2 GHz is obtained. However, for multilayers with bubble state in Py, the permeability spectrum shows an asymmetric broadening peak at 0 magnetic fields. With increasing the magnetic field, the resonance peak also moves to the higher frequency. At higher magnetic field (like larger than 100 Oe), the asymmetry of spectra disappear. The above results indicate that the in-plane uniform precession mode and other resonance modes originated from the non-trivial spin texture of magnetic bubble coexist. The superposition of two kinds of resonance modes broadens the line width.

To analyse the resonance mode carefully, the frequency of in-plane uniform precession mode (Kittel model) versus the magnetic field is fitting, which is shown in Figure 7(a) and 7(e) for bubble state and saturated state, respectively. The main resonance frequency $f_r$ can be determined by the applied field $H_{app}$ according to the Kittel equation [36] as follows,

$$f_r = \frac{\gamma}{2\pi}\sqrt{(4\pi M + H_k + H_{app})(H_k + H_{app})} \qquad (1)$$

Where $\gamma/2\pi=2.94$ GHz/kOe is the gyromagnetic ratio, $4\pi M$ is the magnetization of the films along the applied field $H_{app}$. Generally, $4\pi M$ is much bigger than $H_{app}$ and $H_k$. Thus the above equation (1) can be simplified to be

$$(f_r \times \frac{2\pi}{\gamma})^2 = (4\pi M + H_k) \times H_k + (4\pi M + 2H_k) \times H_{app} \qquad (2)$$

Based on equation (2), the experimental results can be qualitatively fitting [Figure 7(a) and 7(e)]. We can obtain that $4\pi M_{eff}= 9.9$ kGs, $H_k= 4.5$ Oe for bubble states and $4\pi M_{eff}=10.2$ kGs, $H_k=10$ Oe for in-plane saturated state, respectively. Due to the spin non-uniform precession at bubble states, the resonance field $H_k= 4.5$ Oe is smaller than the resonance field $H_k=10.2$ Oe of no bubble states. Since other resonance mode exist besides the kittel mode, so the main peak (peak1) is fitted with Gaussian formula. Except the Kittel peak, another small peak (peak2) is obtained. Figure 7(c) and 7(d) show the resonance absorption peak of bubble states at the field $H_{app}=100$ Oe and $H_{app}=200$ Oe. It can be seen that the linewidth of the main peak becomes narrower and the small peak is not obvious when the field increases. We conclude that the main peak and other mode peak do not move at the same speed with the field. In addition, from the results reported in the literature, we judge that this small peak may correspond to the precession mode of the clockwise (or counterclockwise) of the chiral bubble state. Figure 7(f)-7(h) show resonance absorption peak of in-plane saturated state. The corresponding sweeping fields are 0 Oe, 100 Oe and 200 Oe. One can notice that the single peak Lorentz fitting well corresponding to the experimental results.

## 4.Conclusion

Our results demonstrate that the emergence of metastable bubbles at remnant can be tuned by the tilted field (roughly 3°) along the thin film plane at room temperature in an exchange-coupled Py/[Co/Pd]$_4$ multilayer. Micromagnetic simulation shows that the magnetic moment distribution of magnetic bubbles is same as the Bloch type with the topological number of 1 and -1. Moreover, the high frequency properties of bubble state was studied by VNA-FMR techniques. As the dc magnetic field changes from 0 to 200 Oe, the resonance peak include the Kittel mode and in-plane clockwise or anticlockwise mode of chiral bubble state. We thus provide an effective and promising method for high frequency technological explorations of bubble-based devices.

**Acknowledgments**

This work was supported by the National Natural Science Foundation of China (Nos. 11574121, 51771086).

**Figure captions**

Figure 1. (a) Schematic illustration of exchange-coupled Ta/Pd/[Co/Pd]$_4$/NiFe/Ta multilayers structure. The arrows represent the direction of magnetic moment arrangement. (b) Out-of plane and in-plane hysteresis loop of (Co/Pd)$_4$ multilayers measured by VSM. (c) Out-of-plane and in-plane hysteresis loop of [Co/Pd]$_4$/Py multilayers measured by VSM. The insert graph shows the amplification of in-plane hysteresis loop of [Co/Pd]$_4$/Py multilayers. (d) polar MOKE M–H loops for [Co/Pd]$_4$, Py(15 nm) and [Co/Pd]$_4$/Py(15 nm). (e) transverse MOKE M–H loops for [Co/Pd]$_4$ , Py(15 nm) and [Co/Pd]$_4$/Py(15 nm).

Figure 2. (a)-(d) MFM images for [Co/Pd]$_4$/Py(15 nm) multilayers at different vertical fields. (a) 0 mT, (b) 10 mT, (c) 18 mT, (d) 24 mT. Especially, the red and green circles have no special significance in the Figure 2d. They only represent bubble domains in order to see some clearly. All the scale bar corresponds to 1 $\mu m$.

Figure 3. (a)-(i) MFM imaging of the field-dependent magnetic domain morphology of our [Co/Pd]$_4$/Py(15nm) multilayers at room temperature. The images are captured after magnetized the sample under titled field then removed the field before MFM observation. From 3(a) to 3(i), the values of the magnetic tilted field were respectively: (a) 0 mT, (b)100 mT, (c) 300 mT, (d) 500 mT, (e) 800 mT, (f) 950 mT, (g) -900 mT, (h) -950 mT, (i) 1000 mT. All the scale bar corresponds to 2 $\mu m$.

Figure 4. Bubble stability at room temperature. MFM imaging of domain states when waited a serial of times after removing the magnetic field. (a) 0h, (b) 1h, (c) 3h, (d) 5h, (e) 7h, (f) 9h. All the scale bar corresponds to 2 $\mu m$.

Figure 5. Micromagnetic simulation of (Co/Pd)/Py film. (a)-(h) magnetization configurations under different vertical magnetic field. (a) the initial labyrinth domain states at $B_z$= 0 mT, (b) 50 mT, (c) 100 mT, (d) 150 mT, (e) 200 mT, (f) 250 mT, (g) 300 mT. (h) the magnetic moment distribution of specific magnetic bubbles. (i) the magnetization distributions of the upper Py (left), interface Py (middle) and interface Co(right). All the scale bar corresponds to 200 nm.

Figure 6. The imaginary part of the permeability spectra as a function of frequency under different dc magnetic field for (a) bubble state in Py (b) in-pane saturated state in Py.

Figure 7. Imaginary part of the permeability spectra analysis. (a)-(d) Bubble state (e)-(h) in-plane saturated state. The red solid lines are linear fitting curves. (a) and (e) the applied magnetic field $H_{app}$ dependence on the square of resonance frequency, (b) and (f) DC field $H_{app}$=0 Oe, (c) and (g) $H_{app}$= 100 Oe, (d) and (h) $H_{app}$= 200 Oe.

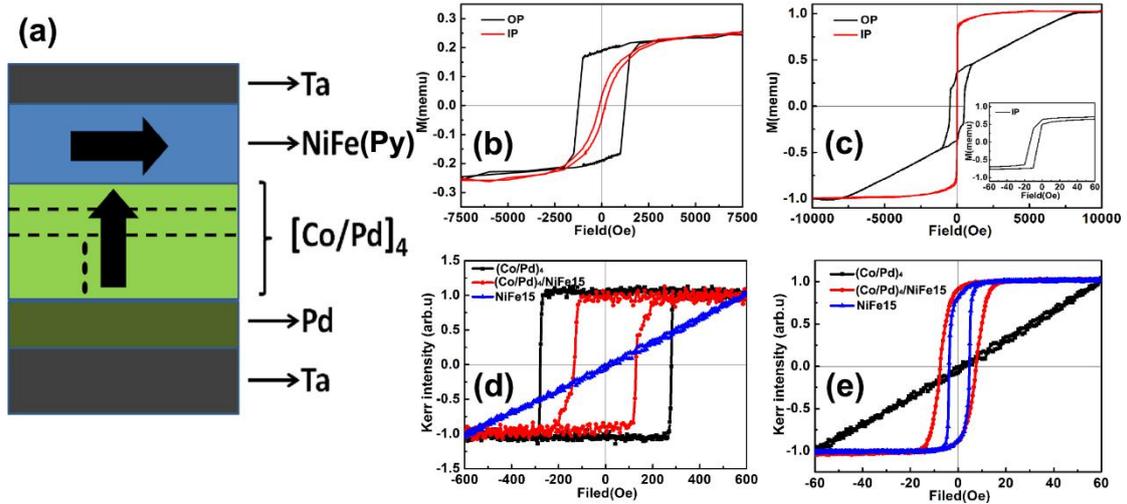

Figure 1

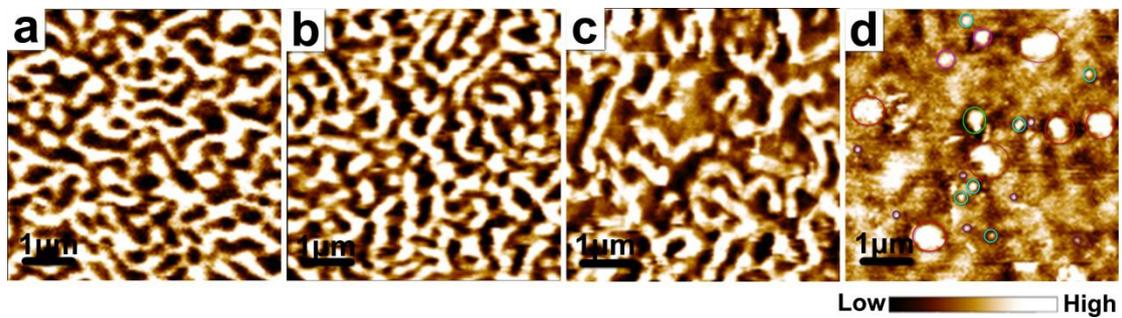

Figure 2

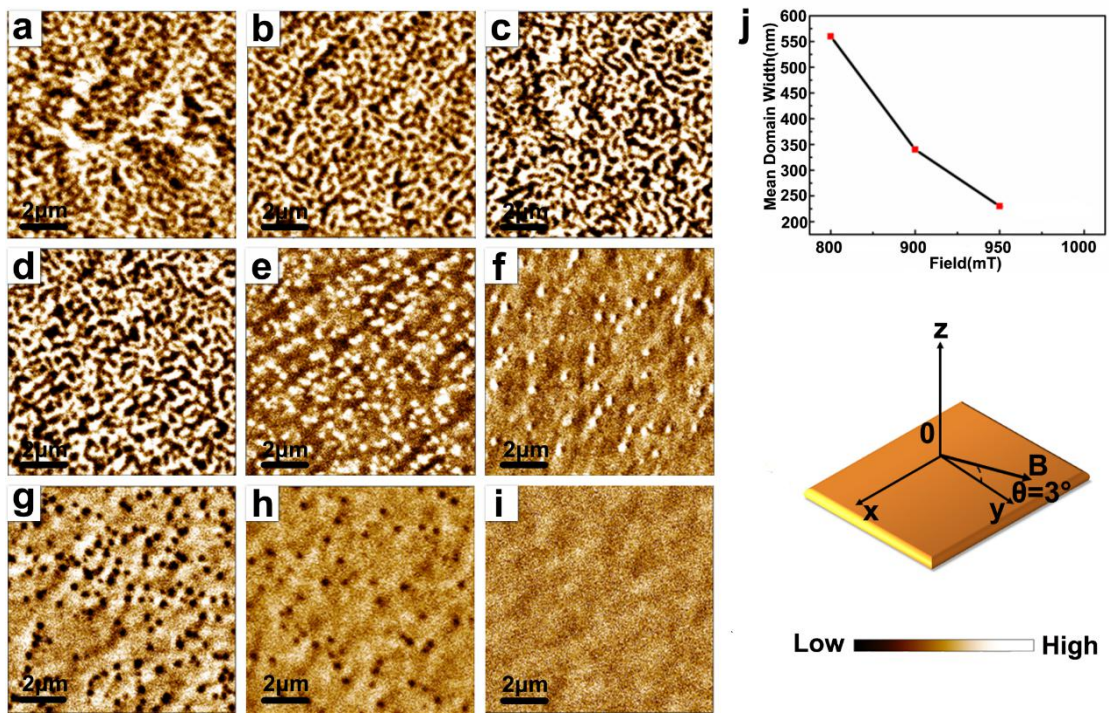

Figure 3

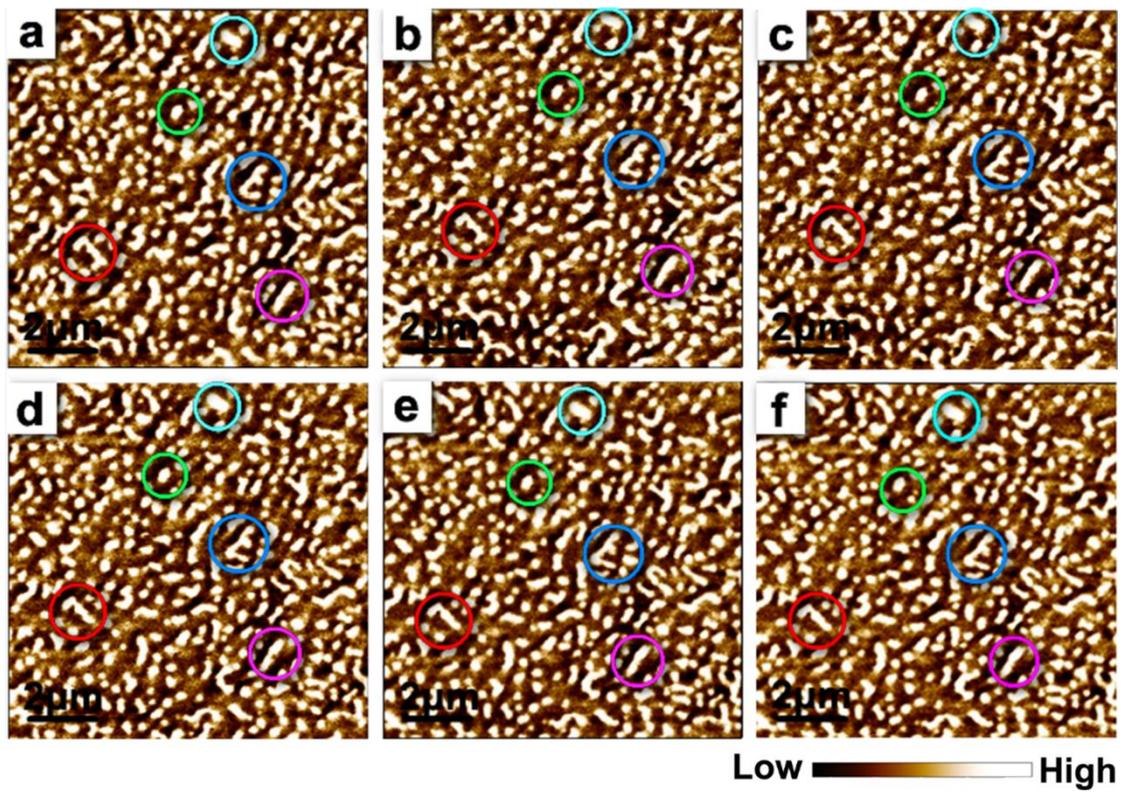

Figure 4

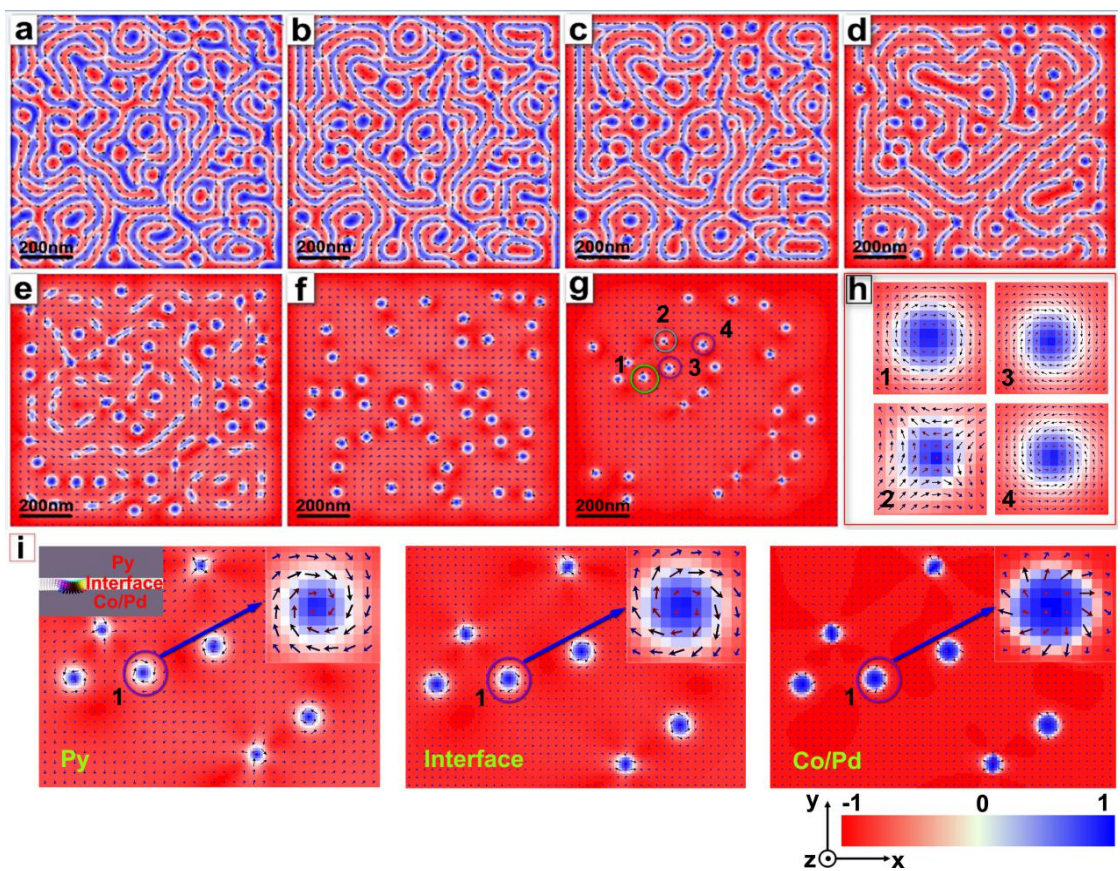

Figure 5

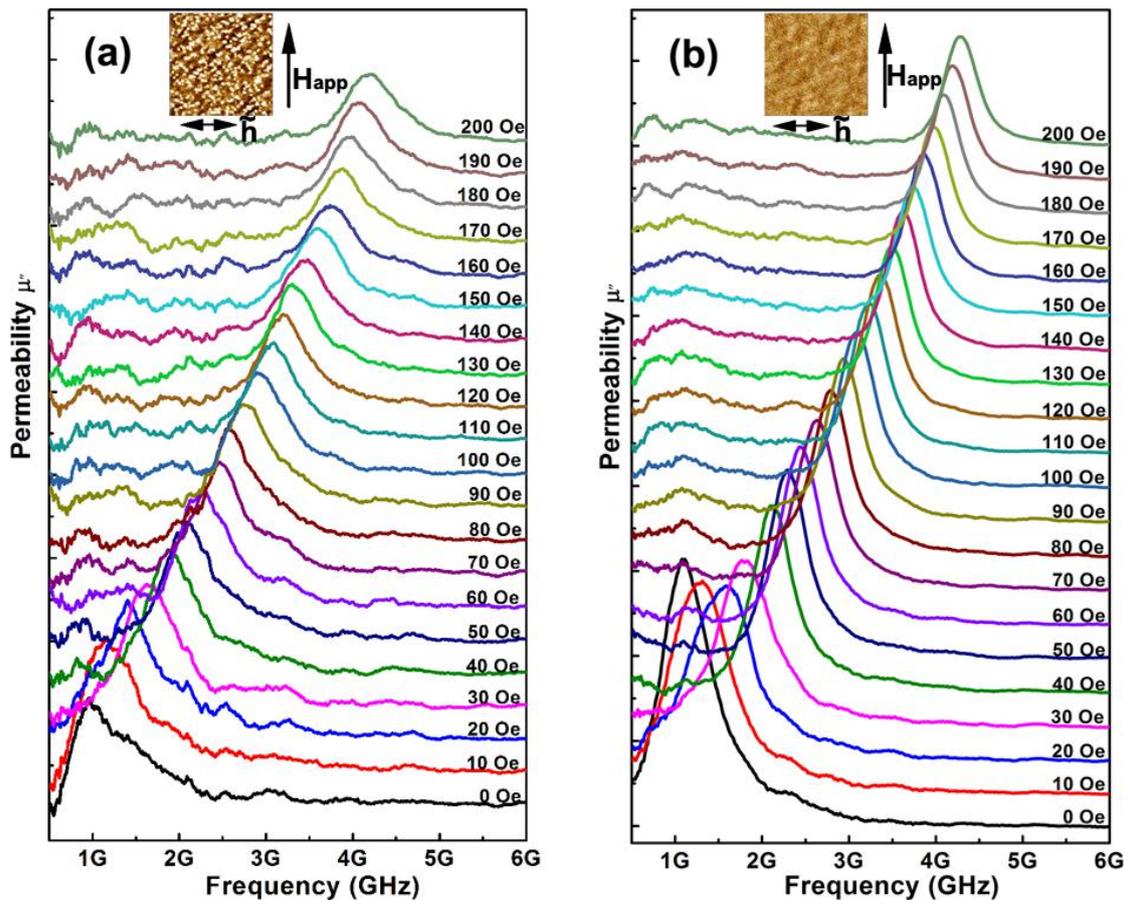

Figure 6

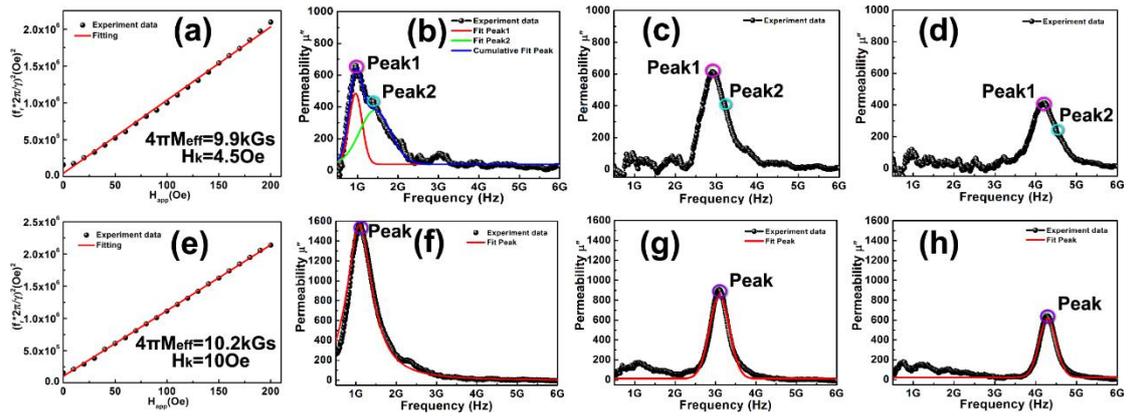

Figure 7